\documentclass[aps,pre,twocolumn,groupedaddress,amsmath,amssymb,preprintnumbers]{revtex4}

\usepackage{graphicx}

\newcommand{\uu}{\mbox{\boldmath$u$}}
\newcommand{\vv}{\mbox{\boldmath$v$}}
\newcommand{\xx}{\mbox{\boldmath$x$}}
\newcommand{\D}{{\mathrm{d}}}
\begin{document}

\title{Stable multispeed lattice Boltzmann methods}
\author{R. A. Brownlee}
\email[corresponding author: ]{r.brownlee@mcs.le.ac.uk}
\author{A. N. Gorban}
\author{J. Levesley}
\affiliation{Department of Mathematics, University of Leicester,
Leicester LE1 7RH, UK}

\date{\today}

\begin{abstract}
    We demonstrate how to produce a stable
    multispeed lattice Boltzmann method (LBM) for
    a wide range of velocity sets, many of which were previously
    thought to be intrinsically unstable. We use
    non-Gauss--Hermitian cubatures. The method operates stably for
    almost zero viscosity, has second-order accuracy,
    suppresses typical spurious oscillation (only a modest Gibbs effect
    is present) and introduces no artificial viscosity. There is
    almost no computational cost for this innovation.

{\bf DISCLAIMER: Additional tests and wide discussion of this
preprint show that the claimed property of coupled steps: no
artificial dissipation and the second-order accuracy of the method
are valid only on sufficiently fine grids. For coarse grids the
higher-order terms destroy coupling of steps and additional
dissipation appears.

The equations are true.}

\end{abstract}

\maketitle

\section{Introduction}\label{sec1}

The lattice Boltzmann method (LBM) is a discrete velocity method
primarily used in the numerical simulation of complex fluid
problems. The essence of the method is that a finite number of
populations stream and collide on a fixed computational lattice in
such a manner that their populations' velocity moments obey the
Navier--Stokes equations.

There are a number of ways to derive the method~\cite{succi01}.
Historically, the method comes from lattice gas automata theory but
it can also be derived directly by the overrelaxation discretization
of Boltzmann's kinetic transport equation.

We prefer to describe the LBM without direct reference to
Boltzmann's equation. Indeed, we prefer to describe the LBM as being
generated wholly by the mechanical motion, of a single-particle
distribution function, by free-flight and entropic involution. In
this setting, the LBM is the discrete dynamical system which arises
by discretizing this motion in a particular manner. The discrete
velocity set arises as approximation nodes of certain cubature in
velocity space~\cite{gorban06}, where these nodes are automorphisms
of some underlying lattice.

A consequence of this realisation is that the stability analysis
(and analysis of conservation laws) of the LBM is more natural. One
is able to clearly identify the main instability mechanisms of the
LBM (Sect.~\ref{sec2}), which are triggered when the continuous
mechanical motion of free-flight and entopic involution is
discretized.

A further consequence of viewing the LBM in this setting (also in
Sect.~\ref{sec2}) is that a prescription of coupled steps is
suggested to stabilise the method. After proceeding for one step of
the usual overrelaxation LBM scheme (LBGK), populations are streamed
and then replaced with their local equilibrium distribution.
Additional dissipation results but the scheme retains the
second-order in time accuracy of LBGK. Compared with the standard
LBM, the proposed scheme of coupled steps constitutes no additional
computational cost.

Each cubature rule gives rise to its own LBM, and multispeed lattice
schemes can be readily concocted (Sect.~\ref{sec3b}). Multispeed
lattices are an attractive prospect because they promise greater
flexibility in the attainable sound speed provided by the model,
allow for the modelling of nonisothermic flows and the dynamics of
higher moments as well. The literature sometimes contains statements
whose sentiment is that some multispeed lattice LBM schemes are
inherently numerically unstable (see,
e.g.,~\cite{dellar05,karlin06_2}). On the contrary, we observe that
the aforementioned LBM of coupled steps is stable for a variety of
multispeed lattices and we demonstrate this through the numerical
simulation of a one-dimensional isothermal shock tube
(Sect.~\ref{sec4}).

\section{The Lattice Boltzmann method}\label{sec2}

To describe the LBM as being generated by free-flight and entropic
involution requires a certain amount of background knowledge and
terminology which we now briefly introduce. For proofs and further
justification of any statements, the reader should
consult~\cite{gorban06,bgl06_2}. The free-flight equation is given
by
\begin{equation}\label{freeflight}
\frac{\partial f}{\partial t}+\vv \cdot \nabla f=0,
\end{equation}
where $f=f(\xx,\vv,t)$ is a single-particle distribution function,
$\xx$ is the space vector, $\vv$ is velocity. This equation
preserves the value of a strictly concave entropy functional,
$S(f)$. The default choice of entropy is
\begin{equation}\label{entropy}
S(f) = -\int f \log{f}\, \D \vv \D \xx.
\end{equation}

In addition, we have a fixed linear mapping $m:f \mapsto M$ to some
macroscopic variables. For example, in hydrodynamic applications $M$
is the vector of five hydrodynamic fields ($n$,$n \uu$,$E$)
(density--momentum--energy):
\begin{equation}\label{hydros}
    n := \int f \D \vv,\quad n u_j := \int v_j f
    \D\vv,\quad E := \frac{1}{2}\int \vv^2 f \D\vv.
\end{equation}

For each $M$, the \textit{quasiequilibrium state} $f^*_M$ is defined
as the unique solution of the constrained optimisation problem:
$\arg\max \{ S(f): m(f)=M \}$. For the hydrodynamic fields
$M=M(\xx,t)$~\eqref{hydros}, the quasiequilibrium state is the well
known local Maxwellian
\begin{equation}\label{maxwell}
  f^*_M(\vv) = n \left( \frac{2\pi k_B T}{m} \right)^{-3/2} \exp{\left(-\frac{m(\vv-\uu)^2}{2 k_B
  T}\right)}.
\end{equation}
where $m$ is particle mass, $k_{\mathrm{B}}$ is Boltzmann's constant
and $T$ is kinetic temperature.

For every $f$, a corresponding quasiequilibrium state $f^*_{m(f)}$
is defined. The set of all quasiequilibrium states is parameterised
by the macroscopic variables, $M$, and defines the
\textit{quasiequilibrium manifold}, which we denote by
$\mathfrak{q}_0$.

The quasiequilibrium approximation for the free-flight
equation~\eqref{freeflight} is the following evolution equation for
the macroscopic variables
\begin{equation}\label{qeapprox}
    \frac{\D M}{\D t}=-m(\vv \cdot \nabla (f^*_M))).
\end{equation}
For the hydrodynamics fields~\eqref{hydros} the
system~\eqref{qeapprox} is the compressible Euler equations.

We denote by $\Pi_S$ the projector of a point $f$ onto
$\mathfrak{q}_0$: $\Pi_S:f \mapsto f^*_{m(f)}$. Let $\Theta_t$ be
the time shift transformation for the free-flight
equation~\eqref{freeflight}: $\Theta_t: f(\xx,\vv) \mapsto f(\xx-\vv
t,\vv)$. Now, for a fixed time step $\tau$, the following step
provides a second-order in time step $\tau$ approximation to the
solution of the conservative macroscopic equations~\eqref{qeapprox}:
\begin{multline}\label{symstep}
M(0) = m(\Pi_S(\Theta_{-\tau/2}(f^*_M))) \\ \mapsto
m(\Pi_S(\Theta_{\tau/2}(f^*_M)))=M(\tau).
\end{multline}
If we would like to model dissipative dynamics then we should follow
the free-flight trajectory by some extra time $\varsigma\leq\tau$:
\begin{multline}\label{combstep}
M(0) = m(\Pi_S(\Theta_{-\vartheta/2}(f^*_M))) \\ \mapsto
m(\Pi_S(\Theta_{\varsigma+\vartheta/2}(f^*_M)))=M(\varsigma+\vartheta)=M(\tau),
\end{multline}
where $\vartheta = \tau-\varsigma$. Now,~\eqref{combstep} provides a
second-order in time step $\tau$ approximation to the solution of
the compressible Navier--Stokes equations with dynamic viscosity
$\mu =\frac{\varsigma}{2} P $ and Prandtl number $Pr=1$.

Equations~\eqref{symstep} and~\eqref{combstep} constitute one step
of an approximation to some conservative and dissipative macroscopic
equations, respectively. To iterate and produce a step wise
approximation to the macroscopic equations for time greater than
$\tau$ we can employ \textit{(partial) entropic involution}.

It will be useful to define the \textit{film of nonequilibrium
states}~\cite{gorban06,Plenka,GorKar} as the manifold $\mathfrak{q}$
that is the trajectory (forward and backward in time) of the
quasiequilibrium manifold. A point $f \in \mathfrak{q}$ is naturally
parameterised by $(M,\tau)$: $f=q_{M,\tau}$, where $M=m(f)$ is the
value of the macroscopic variables, and $\tau = \tau (f)$ is the
time shift from a quasiequilibrium state: $\Theta_{-\tau }(f)$ is a
quasiequilibrium state for some (other) value of $M$. The
quasiequilibrium manifold divides $\mathfrak{q}$ into two parts,
$\mathfrak{q} = \mathfrak{q}_- \cup \mathfrak{q}_0 \cup
\mathfrak{q}_+$, where $\mathfrak{q}_- = \{q_{M, \tau}|\,\tau < 0
\}$ and $\mathfrak{q}_+ = \{q_{M, \tau}|\,\tau > 0 \}$.

Now, the \textit{partial entropic involution} operator $I_S^{\beta}$
is defined as follows: for $f \in \mathfrak{q}_{\pm}$ the point $g =
I_S^{\beta}(f) \in \mathfrak{q}_{\mp}$, for $\beta \in [1/2,1)$, is
specified by the two conditions:
\begin{align*}
m(g)&=m(f),\\
 S(g)-S(\Pi_S(f))&=(2\beta-1)^2(S(f)-S(\Pi_S(f))).
\end{align*}
If $\beta=1$ in this definition we refer to the operator as
\textit{entropic involution} and denote it by $I_S$. The point
$I_S^{\beta}(f)$, $\beta \in [1/2,1)$, is closer to the
quasiequilibrium point $\Pi_S(f)$ (with respect to entropy) than
$I_S(f)$.

If $f_0 \in \mathfrak{q_0}$, then, after some initial steps, the
following sequence gives a second-order in time step $\tau$
approximation of the compressible Navier--Stokes equations (as
mentioned above) with $\varsigma=(1-\beta)\tau/\beta$,
$\beta\in[1/2,1)$:
\begin{align}\label{coup}
M(n\tau) = m((I_S^{\beta} \Theta_{\tau})^n f_0), \quad n=0,1,\ldots.
\end{align}
For entropic involution ($\beta=1$),~\eqref{coup} provides a
second-order in time step $\tau$ approximation of the compressible
Euler equations with time step $2\tau$.

Finally, to generate the standard LBM we must perform three tasks:
\begin{enumerate}
\item transfer to a finite number of velocities with the same macroscopic equations;
\item transfer from space to a lattice, where these velocities are
automorphisms;
\item transfer from dynamics and involution on $\mathfrak{q}$ to the whole space of states.
\end{enumerate}
The first two points will be addressed in Sect.~\ref{sec3a} and
Sect.~\ref{sec3b}. To address the final point we replace the
operator $I_S^{\beta}$ with the transformation
\begin{equation}\label{linInv}
    I_0^{\beta}:f \mapsto \Pi_S(f)+ (2\beta-1)(\Pi_S(f)-f).
\end{equation}
If, for a given $f_0 \in \mathfrak{q_0}$, the sequence~\eqref{coup}
gives the sought after second-order in time step $\tau$
approximation of the macroscopic equations, then the sequence
\begin{equation}\label{coupLin}
    M(n\tau) = m((I_0^{\beta} \Theta_{\tau})^n f_0), \quad
    n=0,1,\ldots,
\end{equation}
also gives a second-order approximation to the same equations.

\subsection{Instability mechanisms}

There are a number of sources of instability which are triggered
when the operator $I_S^\beta$ is replaced with
$I_0^\beta$~\eqref{linInv}. Indeed, we identify three main
instability mechanisms.

\begin{figure}
\begin{centering}
\includegraphics[width=70mm, height=35mm]{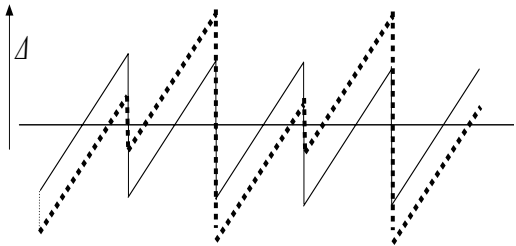}
\caption{\label{NStab}Neutral stability and one-step oscillations in
a sequence successive entropic involutions. Bold dotted line -- a
perturbed motion, $\Delta$ -- direction of neutral
stability.\label{fig1}}
\end{centering}
\end{figure}

Firstly, a small shift in $f$ in the direction of the vector
$f-\Pi_S(f)$ does not relax back for $\beta=1$, and its relaxation
is very slow for $\beta \sim 1$ (i.e., for small viscosity). This
effect is most easily illustrated by Fig.~\ref{NStab}, where a
perturbed chain of entropic involutions is shown. This instability,
which we refer to as \textit{neutral instability}, causes one step
oscillations to be triggered.

Secondly, there is a \textit{nonlinear instability} due to the
nonlinear nature of the projector $\Pi_S$. More specifically, for
all of the accuracy estimates in the previous section, we implicitly
use the assumption that $f$ is sufficiently close to
$\mathfrak{q}_0$. The sequences~\eqref{coup} and~\eqref{coupLin}
give the same accuracy as one another, but a long chain of steps
with the linearised operator $I_0^\beta$~\eqref{linInv} can lead far
from the quasiequilibrium manifold $\mathfrak{q}_0$ and even from
$\mathfrak{q}$ itself.

Finally, there is a \textit{directional instability} that can affect
accuracy. There can be large deviation in the angle the vector
$f-\Pi_S(f)$ makes with the tangent space to $\mathfrak{q}$. The
directional instability changes the structure of dissipation terms
in the target macroscopic equations: accuracy is decreased to the
first-order in $\tau$ and significant fluctuations of the Prandtl
number and viscosity coefficient may occur.

\subsection{Stabilisation}

There is a strikingly simple prescription that simultaneously
alleviates the effect of all three instabilities identified in the
previous section.

\begin{figure}
\begin{centering}
\includegraphics[width=70mm, height=35mm]{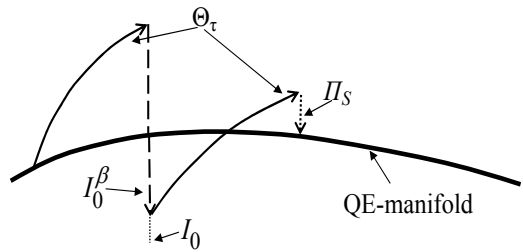}
\caption{\label{CoupledStep}The scheme of coupled steps with
quasiequilibrium ends~\eqref{stabcoupst}.\label{fig2}}
\end{centering}
\end{figure}

Consider starting from a point $f_0 \in \mathfrak{q}_0$, then evolve
the state by $\Theta_{\tau}$ and apply the operator
$I_0^{\beta}$~\eqref{linInv}, as is usual. Then, evolve by
$\Theta_{\tau}$ again and project back on to $\mathfrak{q}_0$ using
$\Pi_S$:
\begin{equation}\label{stabcoupst}
M(0) = m(f_0) \mapsto
m(\Pi_S(\Theta_{\tau}(I_0^{\beta}(\Theta_{\tau}(f_0))))) = M(2\tau).
\end{equation}
This \textit{coupled step} with quasiequilibrium ends is illustrated
in Fig.~\ref{CoupledStep} and gives a second-order in time $\tau$
approximation, to the shift in time $2\tau$, for the target
macroscopic equations with $\varsigma=2(1-\beta)\tau$, $\beta \in
[1/2,1]$. The procedure of periodically restarting from the
quasiequilibrium manifold introduces additional dissipation of order
$\tau^2$, and the perturbation of accuracy is of order $\tau^3$.
Hence, the method has the second-order accuracy.

The user should take into account that $\varsigma$ significantly
depends on the chain construction. For the sequence~\eqref{coup} we
have $\varsigma=(1-\beta)\tau / \beta$, and for the sequence of
steps \eqref{stabcoupst} we have $\varsigma=2(1-\beta)\tau$. The
viscosity coefficient is proportional to $\varsigma$. If $1-\beta$
is small, the coupled step~\eqref{stabcoupst} gives around two times
larger viscosity than~\eqref{coup}.

The result of restarting from $\mathfrak{q}_0$ is that the
aforementioned neutral instability is obliterated and the effects of
nonlinear and directional instabilities are entirely marginalised.

\subsection{Entropy, energy and equilibrium}
\label{sec3a}

Many specific forms of entropy for LBMs have been discussed in the
literature. There exist two methods of construction: from the
Boltzmann entropy approximation to equilibrium (for given
macroscopic variables), and from the equilibrium approximation to
the Boltzmann entropy approximation. For the latter, the universal
entropy formula for a discrete distribution $f=(f_i)$ is the
Kullback entropy
\begin{equation}\label{Massieu-Planck}
S_K(f)=-\sum_{i} f_i \log\biggl(\frac{f_i}{f^*_{m(f), \,i}}\biggr),
\end{equation}
where $f^*_{M}$ is the quasiequilibrium distribution parameterised
by $M$. It is trivial to check that $f^*_{M} = \arg\max \{S_K(f):
m(f)=M \}$. The entropic involution conserves the values of all
elements of $M$. Hence, for energy conservation it is sufficient to
have energy inside $M$ (the distributions $f$ and $f^*_{m(f)}$
should have the same energy). In this sense, the long standing
problem of energy conservation in LBMs has an obvious and physically
meaningful solution (see an example of such a solution
in~\cite{AK2005}). The classical physical sense
of~\eqref{Massieu-Planck} is a nonequilibrium extension of a
Massieu--Planck--Kramers function (or grand canonical potential) for
a given set of macroscopic variables $M$.

A constructive approach to choose discrete $S$ and $f^*_M$ is: we
use the classical (continuous) quasi-equilibrium distributions and
cubature rules to get the appropriate discrete approximation. Then,
the entropy that we need for self-consistency of our construction is
given by~\eqref{Massieu-Planck}.

\section{Cubature and multispeed lattices}\label{sec3b}

To complete the realisation of the LBM, configuration space is
discretized by selecting a finite number of velocities
$\{\vv_1,\vv_2,\ldots,\vv_\ell\}$. More specifically, for a fixed
time step $\tau$, the space discretization will consist of a lattice
$\mathfrak{L}$ for which the velocities $\vv_i \tau$ are
automorphisms. Choosing to discretize in this manner means that no
spatial discretization error is committed when the continuous
free-flight equations
\begin{equation*}
\frac{\partial f_i}{\partial t}+\vv_i \cdot \nabla f=0,\quad
\text{$i=1,\ldots,\ell$},
\end{equation*}
are integrated for time $\tau$ and evaluated on $\mathfrak{L}$.
Here, $f_i$ denotes a single-particle distribution function
associated with the velocity $\vv_i$. Note that all of the
discretization error is contained in the selection of the discrete
velocities.

For hydrodynamics it remains to evaluate the integral
moments~\eqref{hydros}. By construction, the
integrals~\eqref{hydros} remain unchanged under the substitution of
$f$ with the quasiequilibrium state $f^*_M$~\eqref{maxwell}. Our
approach will be to construct cubature rules based upon reproduction
of low degree polynomials using the discrete velocities, $\vv_i$, as
abscissas. We replace the quasiequilibrium state~\eqref{maxwell}
with its second-order Taylor expansion in $\uu$ with terms involving
$\uu^2$ and higher disregarded. We denote this polynomial by
$p^*_M=p^*_M(\vv)$, and demand that its first few moments are
evaluated exactly by the cubature rule. Another approach, which we
do not pursue here, is to employ a discrete version of the
entropy~\eqref{entropy} from the outset (see,
e.g.,~\cite{karlin02}).

Let us concern ourselves with modelling the isothermal
Navier--Stokes equations for the remainder of the paper and set
$C^2= 2 k_B T/m$. The ``sound speed'' of the model is $c_s$ where
$c_s^2=C^2/2$. For the isothermal model we should at least consider
all moments up to the third-order, which is equivalent to finding
the cubature weights $w_i$ such that
\begin{equation*}
    \int \vv^k \exp{\Bigl(-\frac{\vv^2}{C^2}\Bigr)}\,\D\vv \equiv \sum_{i=1}^\ell w_i \vv_i^k
    \exp{\Bigl(-\frac{\vv^2_i}{C^2}\Bigr)},
\end{equation*}
for $k=0,1,2,3$. Using this cubature, the discrete hydrodynamic
moments are given by the expressions
\begin{equation*}
    n = \sum_i f_i,\quad n \uu = \sum_i \vv_i f_i,\quad E= \frac{1}{2}\sum_i \vv^2_i
    f_i,
\end{equation*}
where $f_i=f_i(\xx,t):=w_i f(\xx,\vv_i,t)$. The LBM realisation
of~\eqref{coup} is then
\begin{equation}\label{LBGK}
     f_i(\xx+\vv_i\tau,t+\tau) = f_i^*+(2\beta-1)(f_i^*-f_i),
\end{equation}
where $f_i^*=f_i^*(\xx,t):=w_i p^*_M(\vv_i)$. The name often given
to~\eqref{LBGK} in the literature is LBGK. The prescribed scheme of
coupled steps~\eqref{stabcoupst} is:
\begin{equation}\label{coupLB}
     f_i(\xx+\vv_i\tau,t+\tau) = \left\{
    \begin{aligned}
        &f_i^*,&& \text{$N_{\mathrm{step}}$ odd,} \\
        &f_i^*+(2\beta-1)(f_i^*-f_i),&& \text{otherwise,}
    \end{aligned}\right.
\end{equation}
where $N_{\mathrm{step}}$ is the cumulative number of time steps
taken in the simulation.

Clearly, each set $(\vv_i,w_i)$ of velocities and associated
cubature weights will define its own LBM.

One way to handle the multivariate case is to use tensor and
algebraic products of one-dimensional velocities and weights,
respectively. Let us therefore restrict our discussion to one space
dimension.

One particular choice of quadrature is Gauss--Hermite quadrature.
Indeed, the three-point Gauss--Hermite rule (which exactly evaluates
the fifth order moment) is popular~\cite{luo97,he98}. However, the
use of multispeed lattices (lattices with more then two non-zero
speeds in one-dimension) using higher-order Gauss--Hermite
quadrature is precluded because the zeros of the Hermite
polynomials, for degree greater than $3$, can not be aligned with
any regular lattice.

An alternative to Gauss--Hermite is to use Newton--Cotes quadrature
which has the advantage that the velocities can be aligned with a
lattice before determining the quadrature weights. Multispeed
lattices can be readily constructed. For example, suppose we want to
employ $5$ symmetric quadrature nodes
$\{\vv_0,\ldots,\vv_4\}:=\{0,\pm a C,\pm b C\}$. Then, $w_i = C W_i
\sqrt{\pi} \exp{(\vv^2_i/C^2)} $, where
\begin{align*}
   W_0 &= \frac{4a^2b^2-2 a^2-2b^2+3}{4a^2b^2},\\
   W_1 &= W_2 =  \frac{2b^2-3}{8(b^2-a^2)a^2},\quad W_3 = W_4 =  \frac{2a^2-3}{8(a^2-b^2)b^2}.
\end{align*}
For five distinct nodes we should have $0<a<b$ but there should be
further restrictions on $a$ and $b$ to ensure all weights are
positive.

A selection of different quadrature rules are collected in
Table~\ref{tab1} along with their respective lattice ``sound
speeds''.

\begin{table}
\centering
\begin{tabular}{ccccc}
\hline\hline
Method & Order & $\vv_i$ & $W_i$ & $c_s$\\
\hline G--H & 5 & $\{ 0,\pm1 \}$ & $\{\frac{2}{3},\frac{1}{6}\}$ & $1/\sqrt{3} $ \\
N--C & 4 & $\{0,\pm1,\pm2 \}$ & $\{\frac{9}{16},\frac{5}{24},\frac{1}{96} \}$ & $1/\sqrt{2}$ \\
N--C & 4 & $\{0,\pm1,\pm2 \}$ & $\{\frac{1}{2},\frac{1}{6},\frac{1}{12} \}$ & $1$ \\
N--C & 4 & $\{0,\pm1,\pm 3 \}$ & $\{\frac{19}{36},\frac{15}{64},\frac{1}{576} \}$ & $1/\sqrt{2}$ \\
N--C & 4 & $\{0,\pm1,\pm 3 \}$ & $\{\frac{2}{9},\frac{3}{8},\frac{1}{72} \}$ & $1$ \\
N--C & 4 & $\{0,\pm1,\pm 4 \}$ & $\{\frac{33}{64},\frac{29}{120},\frac{1}{1920} \}$ & $1/\sqrt{2}$ \\
N--C & 4 & $\{0,\pm1,\pm 4 \}$ & $\{\frac{1}{8},\frac{13}{30},\frac{1}{240} \}$ & $1$ \\
N--C & 6 & $\{0,\pm1,\pm2,\pm3 \}$ & $\{\frac{161}{288},\frac{27}{128},\frac{3}{320},\frac{1}{5760} \}$ & $1/\sqrt{2}$  \\
N--C & 6 & $\{0,\pm1,\pm2,\pm3 \}$ & $\{\frac{7}{18},\frac{1}{4},\frac{1}{20},\frac{1}{180} \}$ & $1$  \\
\hline\hline
\end{tabular}
\caption{Various 1D lattices and weights using different
quadratures: G--H means Gauss--Hermite, N--C means Newton--Cotes.
The second column shows the order of the moment which is exactly
evaluated by the quadrature rule. The final column gives the lattice
``sound speed''.\label{tab1}}
\end{table}

\section{Numerical experiment}\label{sec4}

The $1$D shock tube for a compressible isothermal fluid is a
standard benchmark test for hydrodynamic codes. We will fix the
kinematic viscosity of the fluid at $\nu=10^{-9}$. Our computational
domain will be the interval $[0,1]$ and we discretize this interval
with $801$ uniformly spaced lattice sites. We choose the initial
density ratio as 1:2 so that for $x\leq 400$ we set $n=1.0$ else we
set $n=0.5$.

\begin{figure}
\begin{centering}
\includegraphics[width=85mm]{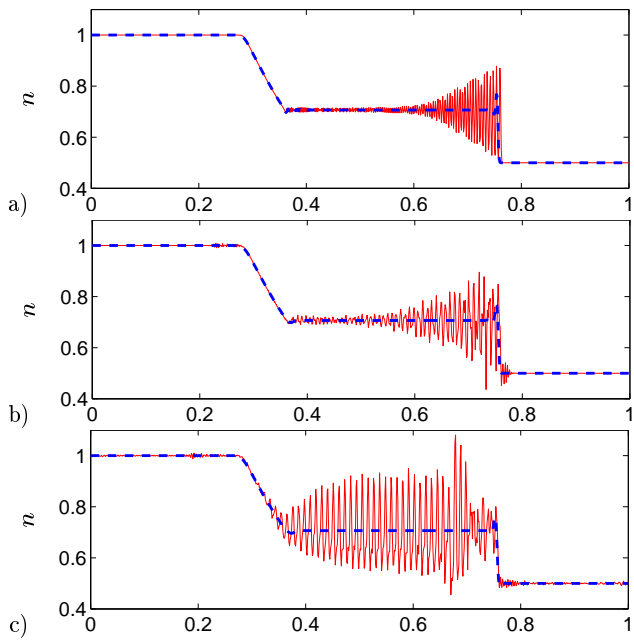}
\caption{Density profile of a 1D isothermal shock tube simulation
after $100\sqrt{3}/c_s$ time steps using LBGK (solid line) and the
coupled steps~\eqref{coupLB} (dashed line) with the lattice a)
$(\vv_i,W_i,c_s)=(\{0,\pm1\},\{\frac{2}{3},\frac{1}{6}\},1/\sqrt{3})$;
b)
$(\vv_i,W_i,c_s)=(\{0,\pm1,\pm2\},\{\frac{9}{16},\frac{5}{24},\frac{1}{96}
\},1/\sqrt{2})$; c)
$(\vv_i,W_i,c_s)=(\{0,\pm1,\pm2,\pm3\},\{\frac{7}{18},\frac{1}{4},\frac{1}{20},\frac{1}{180}
\},1)$ }\label{fig3}
\end{centering}
\end{figure}

In all of our simulations we use a lattice with unit spacing, a unit
time step and we have chosen to consider the three-velocity
Gauss-Hermite model, as well as five- and seven-velocity
Newton-Cotes models (see Fig.~\ref{fig3} for the specific lattices
details).

The standard three-velocity LBGK discretization~\eqref{LBGK} is
violently oscillatory in the immediate neighbourhood of the shock,
the effects of which extend over much of the domain. The situation
is worse for LBGK with five- and seven-velocities; here the
oscillations do not remain bounded (we always implement the
positivity rule~\cite{bgl06_01,bgl06_2} that prohibits negative
densities). The scheme quickly becomes meaningless. In comparison,
we find that the proposed scheme of coupled steps~\eqref{coupLB}
gives stable behaviour on all tested lattices, by which we mean no
blow-up and spurious post-shock oscillations are observed. In all
cases we do observe a small deviation near the shock front which may
be an unavoidable Gibbs effect. The amplitude of the Gibbs effect in
our experiments decreases by increasing the degree of approximation.

\section{Conclusions}

By considering a realisation in which the LBM is wholly generated by
free-flight and entropic involution, we have suggested a simple
stabilisation procedure of coupled steps~\eqref{coup} which retains
the second-order accuracy of the method without artificial viscosity
at that order.

The procedure effectively constitutes no additional computational
cost and can be implemented in any existing LBM code with just a few
changes.

We have demonstrated that the oscillatory pattern of the LBM in the
vicinity of shocks is not due to a lack of artificial diffusivity
and are not pertinent to proper lattice Boltzmann schemes. On the
other hand, there exists a Gibbs effect that we cannot fully
suppress without artificial viscosity or special ad hoc
approximation schemes.

The role of Gauss--Hermite quadratures in LBM constructions is now
overestimated (here we fully agree with~\cite{karlin06_2}). Other
quadratures require slightly more points for the same degree of
accuracy, but can be realised on integer nodes (and, more generally,
are flexible with regards to the choice of nodes), which is more
convenient for LBM needs.

The five-velocity lattice $\{0,\pm 1,\pm 2 \}$ can produce a stable
LBGK based method, as can many other ``suspicious" lattices which
were previously discussed by many authors as intrinsically unstable.
These results can be immediately generalised onto high dimensional
lattices ($N$D-lattices, cubatures and equilibria as products of 1D
lattices, quadratures and equilibria). Finally, with the advent of
stable multispeed lattice formulations comes the prospect of stable
nonisothermic LBM realisations.

\section*{Acknowledgements}

This work is supported by Engineering and Physical Sciences Research
Council (EPSRC) grant number GR/S95572/01.

\bibliographystyle{apsrev}


\begin{thebibliography}{10}
\expandafter\ifx\csname
natexlab\endcsname\relax\def\natexlab#1{#1}\fi
\expandafter\ifx\csname bibnamefont\endcsname\relax
  \def\bibnamefont#1{#1}\fi
\expandafter\ifx\csname bibfnamefont\endcsname\relax
  \def\bibfnamefont#1{#1}\fi
\expandafter\ifx\csname citenamefont\endcsname\relax
  \def\citenamefont#1{#1}\fi
\expandafter\ifx\csname url\endcsname\relax
  \def\url#1{\texttt{#1}}\fi
\expandafter\ifx\csname urlprefix\endcsname\relax\def\urlprefix{URL
}\fi \providecommand{\bibinfo}[2]{#2}
\providecommand{\eprint}[2][]{\url{#2}}

\bibitem[{\citenamefont{Succi}(2001)}]{succi01}
\bibinfo{author}{\bibfnamefont{S.}~\bibnamefont{Succi}},
  \emph{\bibinfo{title}{The lattice {B}oltzmann equation for fluid dynamics and
  beyond}} (\bibinfo{publisher}{OUP}, \bibinfo{address}{New York},
  \bibinfo{year}{2001}).

\bibitem[{\citenamefont{Gorban}(2006)}]{gorban06}
\bibinfo{author}{\bibfnamefont{A.~N.} \bibnamefont{Gorban}}, in
  \emph{\bibinfo{booktitle}{Model Reduction and Coarse-Graining Approaches for
  Multiscale Phenomena}} (\bibinfo{publisher}{Springer},
  \bibinfo{address}{Berlin-Heidelberg-New York}, \bibinfo{year}{2006}), pp.
  \bibinfo{pages}{117--176}, \bibinfo{note}{cond-mat/0602024}.

\bibitem[{\citenamefont{Dellar}(2005)}]{dellar05}
\bibinfo{author}{\bibfnamefont{P.~J.} \bibnamefont{Dellar}}, in
  \emph{\bibinfo{booktitle}{Computational Fluid and Solid Mechanics 2005}}
  (\bibinfo{publisher}{Elsevier}, \bibinfo{address}{Amsterdam},
  \bibinfo{year}{2005}), pp. \bibinfo{pages}{632--635}.

\bibitem[{\citenamefont{Chikatamarla and Karlin}(2006)}]{karlin06_2}
\bibinfo{author}{\bibfnamefont{S.~S.} \bibnamefont{Chikatamarla}}
  \bibnamefont{and} \bibinfo{author}{\bibfnamefont{I.~V.}
  \bibnamefont{Karlin}}, \bibinfo{journal}{Phys. Rev. Lett.}
  \textbf{\bibinfo{volume}{97}}, \bibinfo{pages}{190601}
  (\bibinfo{year}{2006}).

\bibitem[{\citenamefont{Brownlee et~al.}(2006)\citenamefont{Brownlee, Gorban,
  and Levesley}}]{bgl06_01}
\bibinfo{author}{\bibfnamefont{R.~A.} \bibnamefont{Brownlee}},
  \bibinfo{author}{\bibfnamefont{A.~N.} \bibnamefont{Gorban}},
  \bibnamefont{and} \bibinfo{author}{\bibfnamefont{J.}~\bibnamefont{Levesley}},
  \bibinfo{journal}{Phys. Rev. E }
  \textbf{\bibinfo{volume}{74}}, \bibinfo{pages}{037703}
  (\bibinfo{year}{2006}).

\bibitem[{\citenamefont{Brownlee et~al.}(2006)\citenamefont{Brownlee, Gorban,
  and Levesley}}]{bgl06_2}
\bibinfo{author}{\bibfnamefont{R.~A.} \bibnamefont{Brownlee}},
  \bibinfo{author}{\bibfnamefont{A.~N.} \bibnamefont{Gorban}},
  \bibnamefont{and} \bibinfo{author}{\bibfnamefont{J.}~\bibnamefont{Levesley}},
  \bibinfo{journal}{Phys. Rev. E (submitted)}  (\bibinfo{year}{2006}),
  \bibinfo{note}{cond-mat/0611444}.

\bibitem[{\citenamefont{Gorban and Karlin}(2003)}]{Plenka}
\bibinfo{author}{\bibfnamefont{A.~N.} \bibnamefont{Gorban}} \bibnamefont{and}
  \bibinfo{author}{\bibfnamefont{I.~V.} \bibnamefont{Karlin}},
  \bibinfo{type}{Preprint} \bibinfo{number}{IHES/P/03/57},
  \bibinfo{institution}{Institut des Hautes \'Etudes Scientifiques},
  \bibinfo{address}{Bures-sur-Yvette, France} (\bibinfo{year}{2003}),
  \bibinfo{note}{cond-mat/0308331}.

\bibitem[{\citenamefont{Gorban and Karlin}(2005)}]{GorKar}
\bibinfo{author}{\bibfnamefont{A.~N.} \bibnamefont{Gorban}} \bibnamefont{and}
  \bibinfo{author}{\bibfnamefont{I.~V.} \bibnamefont{Karlin}},
  \emph{\bibinfo{title}{Invariant manifolds for physical and chemical
  kinetics}}, vol. \bibinfo{volume}{660} of \emph{\bibinfo{series}{Lect. Notes
  Phys.}} (\bibinfo{publisher}{Springer},
  \bibinfo{address}{Berlin-Heidelberg-New York}, \bibinfo{year}{2005}).

\bibitem[{\citenamefont{Ansumali and Karlin}(2002)}]{karlin02}
\bibinfo{author}{\bibfnamefont{S.}~\bibnamefont{Ansumali}} \bibnamefont{and}
  \bibinfo{author}{\bibfnamefont{I.~V.} \bibnamefont{Karlin}},
  \bibinfo{journal}{J. Stat. Phys.} \textbf{\bibinfo{volume}{107}},
  \bibinfo{pages}{291} (\bibinfo{year}{2002}).

\bibitem[{\citenamefont{Ansumali and Karlin}(2005)}]{AK2005}
\bibinfo{author}{\bibfnamefont{S.}~\bibnamefont{Ansumali}} \bibnamefont{and}
  \bibinfo{author}{\bibfnamefont{I.~V.} \bibnamefont{Karlin}},
  \bibinfo{journal}{Phys. Rev. Lett.} \textbf{\bibinfo{volume}{95}},
  \bibinfo{pages}{260605} (\bibinfo{year}{2005}).


\bibitem[{\citenamefont{He and Luo}(1997)}]{luo97}
\bibinfo{author}{\bibfnamefont{X.}~\bibnamefont{He}} \bibnamefont{and}
  \bibinfo{author}{\bibfnamefont{L.-S.} \bibnamefont{Luo}},
  \bibinfo{journal}{Phys. Rev. E} \textbf{\bibinfo{volume}{55}},
  \bibinfo{pages}{R6333} (\bibinfo{year}{1997}).

\bibitem[{\citenamefont{Shan and He}(1998)}]{he98}
\bibinfo{author}{\bibfnamefont{X.}~\bibnamefont{Shan}} \bibnamefont{and}
  \bibinfo{author}{\bibfnamefont{X.}~\bibnamefont{He}}, \bibinfo{journal}{Phys.
  Rev. Lett.} \textbf{\bibinfo{volume}{80}}, \bibinfo{pages}{65}
  (\bibinfo{year}{1998}).

\end{thebibliography}

\end{document}